\documentclass[prb,twocolumn,superscriptaddress]{revtex4}
\usepackage{graphicx}  
\usepackage[colorlinks=true]{hyperref}
\usepackage[dvipsnames]{xcolor}

\newcommand{\zSe}{\ensuremath{\mathrm{z_{Se} \ }}}

\newcommand{\Gm}{\ensuremath{\mathrm{\Gamma \ }}}

\newcommand{\Ag}{\ensuremath{\mathrm{A_{1g} \ }}}
\newcommand{\Au}{\ensuremath{\mathrm{A_{2u} \ }}}
\newcommand{\Bg}{\ensuremath{\mathrm{B_{1g} \ }}}
\newcommand{\Eg}{\ensuremath{\mathrm{E_{g} \ }}}
\newcommand{\Eu}{\ensuremath{\mathrm{E_{u} \ }}}

\begin{document}

\title{Correlation Driven Phonon Anomalies in Bulk FeSe}
\author{Ghanashyam Khanal}
\affiliation{Department of Physics \& Astronomy, Rutgers University, Piscataway, NJ 08854-8019, USA}
\author{Kristjan Haule}
\affiliation{Department of Physics \& Astronomy, Rutgers University, Piscataway, NJ 08854-8019, USA}

\date{\today}

\begin{abstract}
We study the lattice dynamics of iron superconductor FeSe, and address the fundamental question of how important is proper description of fluctuating magnetic moments in metallic systems for phonon dispersion and phonon density of states. We show that Density Functional Theory (DFT)+ embedded Dynamical Mean-Field Theory (eDMFT) functional approach, which  truly captures the fluctuating local moments, largely eliminates the deficiency of DFT for description of lattice dynamics in correlated metallic systems, and predicts phonon dispersion and phonon density of states in very good agreement with available X-ray data and nuclear inelastic scattering. This benchmark between eDMFT and experiment will be important for  data science driven material design, in which DFT is being replaced by beyond DFT methods.
\end{abstract}
\maketitle

{\color{NavyBlue}\textit{Introduction:}} The unconventional superconductivity discovered in iron-based superconductors (FeSC) more than a decade ago~\cite{Kamihara_2008} has been mostly interpreted in terms of an unconventional pairing due to electron-electron interactions~\cite{Dai2012,YinSC,RevModPhys.83.1589}, and spin-fluctuations mechanism~\cite{spin_fluctuations,RevModPhys.84.1383}. However, it was also pointed out that the electron-phonon coupling (EPC) in these superconductors is substantial~\cite{Haule_FeSe_2014,Gerber_Science_2017,Kim2012}, and plays an important role in boosting superconducting temperature~\cite{Haule_FeSe_2014,Gerber_Science_2017,Kim2012,Coh_2015,PhysRevB.93.134513}, as well as boosting nematicity scale in FeSe.
FeSe is one of the simplest FeSC with the superconducting T$_c$ of 8$\,$K in the bulk~\cite{Hsu_2008}, which is increased to 37$\,$K under pressure~\cite{Medvedev_2009,Okabe_2010}, and when the monolayer FeSe is grown on top of SrTiO$_3$, the transition temperature increases further to 65$\,$K~\cite{He_2013}.

Theoretically, FeSe is one of the most correlated FeSC~\cite{Yin2011} in particular the $xy$ orbital shows very strong mass enhancement and incoherence~\cite{Yin2011, Yi2015,Aichhorn_2013, Mandal_FeSe_2017}. Consequently FeSe was categorized as one of the best examples of Hund's metals~\cite{Haule_njp} with strongest orbital differentiation~\cite{Yin2011}, sometimes called orbital selectivity~\cite{QSiOSP,Yi2015}. Namely, in these systems the Hund's coupling slows down the electrons, such that low temperature quasiparticles have substantial mass enhancement~\cite{Yin2011}, and different orbitals show different correlation strength and different coherence scale, hence the appearance of orbital differentiation~\cite{Yin2011,Yin2016}.
At finite temperature, some orbitals that have coherence scale larger than the temperature of measurement appear metallic, and others appear incoherent at the same temperature. Consequently, the temperature at which orbital fluctuations are arrested is much higher than the temperature at which spin fluctuations are screened, and the latter are responsible for low energy and low temperature anomalous properties. 

The chalcogen height and/or bond angle between Se-Fe-Se plays an important role in determining the strength of the correlations~\cite{Yin2014,ct28,Yin2011}, magnetic excitation spectra~\cite{Yin2014}, and T$_c$~\cite{PhysRevB.81.205119}. Conversely correlations due to Hund's coupling substantially increase the chalcogen height, as compared to Density Functional Theory (DFT) prediction, they also substantially enhance the electron-phonon coupling, and make the  A1g phonon mode, associated with chalcogen height, softer than appears in DFT, as was predicted in Ref.~\onlinecite{Haule_FeSe_2014} by embedded Dynamical Mean Field Theory (eDMFT) method. This was later confirmed by pioneering measurement of the electron-phonon coupling, utilizing photoemission spectroscopy in lock with x-ray diffraction.~\cite{Gerber_Science_2017}.
While eDMFT predicts very accurate crystal structure of FeSe~\cite{Haule_FeSe_2014,Haule_2016_Force}, it was pointed out quite early on~\cite{SeheightDFT, SeheightDFT2, SeheightDFT3, SeheightDFT4} that simpler DFT calculation in magnetically ordered state can give almost as good crystal structure as eDMFT. Hence static magnetic moments give similar crystal structure as fluctuating moments. However, the electronic structure and the Fermi surface of the magnetic DFT is very different from the measured photoemission data~\cite{Photoemission}, which is in much better agreement with non-magnetic DFT, and in even better agreement with DMFT~\cite{Aichhorn_2013, Photoemission}. Hence, in order to properly predict the structural degrees of freedom one needs magnetism in DFT, while prediction of electronic band-structure requires absence of magnetism. Several DFT-based works studied phonon spectra in the presence and absence of long range magnetism~\cite{Subedi_2008, Nakamura_2009, Cohen_2012, Xing_2014, Koufos_2014, Zakeri_2017}, and tried to answer the question whether
phonon spectra is better predicted in the magnetic or non-magnetic state. However, both results significantly deviate from measured phonon-spectra~\cite{Murai_2020,Cava_2010}.
Here we want to answer more fundamental question, namely, can methods that rely on static magnetic moments mimic fluctuating moments for the purpose of predicting phonon spectra? To address this question, one needs a method that truly captures fluctuating local moments, like the eDMFT method.

The correlations that allow local moments on iron to coexist with metallic bands, make the iron atom larger than it would be if electrons were very itinerant~\cite{Haule_FeSe_2014}. The larger Fe atoms push Se away from the Fe-plane, and consequently the Fe-Se distance increases, which therefore reduces the hybridization between Fe and Se, and increases the correlation strength on Fe even further, because smaller hybridization makes screening of local moments less efficient. Consequently, this positive feedback loop makes A1g phonon frequency very strongly coupled with degree of correlations on Fe atom, and the phonon mode is much softer in eDMFT than in DFT~\cite{Haule_FeSe_2014, Gerber_Science_2017}. 
The important question that we would like to address in this work is how are the other phonon frequencies affected by correlations, and whether predominantly Fe or Se modes are affected more by the presence of fluctuating moments. Finally, we will address the difference between the static versus fluctuating moments and its difference with the itinerant electrons.

{\color{NavyBlue}\textit{Computational Details:}} In this study we use stationary and charge self consistent implementation of DFT+eDMFT~\cite{Haule_DMFT_2010, JPSJ_Haule,Haule_webpage}, which properly captures local spin fluctuations, and has previously been shown to predict well the electronic properties of FeSC~\cite{Yin-np11,Yin2011,PhysRevB.86.195141,Wang2013,Yin2014,YinSC,PhysRevLett.116.247001,PhysRevB.96.195121,PhysRevLett.119.096401,PhysRevB.90.060501,Haule_FeSe_2014,Mandal_FeSe_2017}. The forces on atoms are calculated by the analytical derivative of the free energy functional with respect to the displacement of each atom~\cite{Haule_Free_Energy_2015, Haule_2016_Force}, hence the electronic entropy of a spin fluctuating moments is included in the free energy and the force, as opposed to the DFT zero temperature method. This has significant impact on stabilizing phonon spectra, as shown for paramagnetic bcc phase of elemental iron in Ref.~\citenum{QHan}.
Continuous time quantum Monte Carlo (CTQMC)~\cite{Haule_CTQMC_2007} was used as the impurity solver for the quantum impurity problem, in which the Coulomb interaction ($U$) and Hund's exchange interaction ($J$) take the value $5.0\,$eV and $0.8\,$eV, as previously established in numerous works~\cite{Yin-np11,Yin2011,PhysRevB.86.195141,Wang2013,Yin2014,YinSC,PhysRevLett.116.247001,PhysRevB.96.195121,PhysRevLett.119.096401,PhysRevB.90.060501,Haule_FeSe_2014,Mandal_FeSe_2017}. As in these previous works, the nominal double counting was used, which was shown to be very close to exact double counting~\cite{Haule_DC_2015}. The DFT Kohn-Sham orbitals are computed within the Wien2k package~\cite{Wien2k_2001}, and we use the Local Density Approximation (LDA) functional for the exchange correlation part in DFT part of DFT+eDMFT. All calculations are performed at $T=116 K$. 

FeSe crystallizes in tetragonal structure (space group P4/nmm, no. 139) with the lattice constants $a=3.7685\,\textrm{\AA} \text{ and } c=5.5194\,\textrm{\AA}$. The Fe-layers share the edge with Se-layers at a distance \zSe above and below the Fe-plane. This internal parameter (\zSe) was optimized within eDMFT method using a mesh of $2000$ k-points in the full BZ. The Se atomic position was relaxed until the forces on atoms were smaller than 0.001 eV/$\textrm{\AA}$.
For the phonon calculation the $2\times 2\times 2$ supercell was constructed with the atoms displaced from their equilibrium position, as allowed by the symmetry of the system. Once the forces on each atom are calculated within the eDMFT framework~\cite{Haule_2016_Force}, the phonon dispersion and the phonon density of states are calculated using the first principle finite displacement method, as implemented in Phonopy package \cite{Togo_2015_phonopy}. For the comparison with previous DFT calculations we use Generalized Gradient Approximation (GGA) with PBE functional as DFT method, to be consistent with previous DFT works.

{\color{NavyBlue}\textit{Results:}} The optimization of the Se height \zSe is not only essential for calculating the phonon properties, but it also determines the correlation strength~\cite{Haule_FeSe_2014}, and was even linked to the value of $T_c$~\cite{Okabe_2010}. Because of the overbinding between Fe and Se, the LDA method predicts \zSe=0.232 wheres  GGA gives \zSe=0.243, which are both well below the experimental value of 0.265-0.268~\cite{McQueen_2009,Gerber_Science_2017}. The correlations on iron make the iron ion larger, and hence \zSe increases to value of 0.270 within eDMFT, as was previously reported in Refs.~\citenum{Haule_FeSe_2014,Haule_2016_Force}. This is extremely close to the experimental value from Gerber \textit{et.al.}~\cite{Gerber_Science_2017} of 0.268, hence we expect this remarkable improvement of the structure going from DFT to eDMFT will have significant impact on vibrational properties.

\begin{figure}[t]
	\includegraphics[width=\columnwidth]{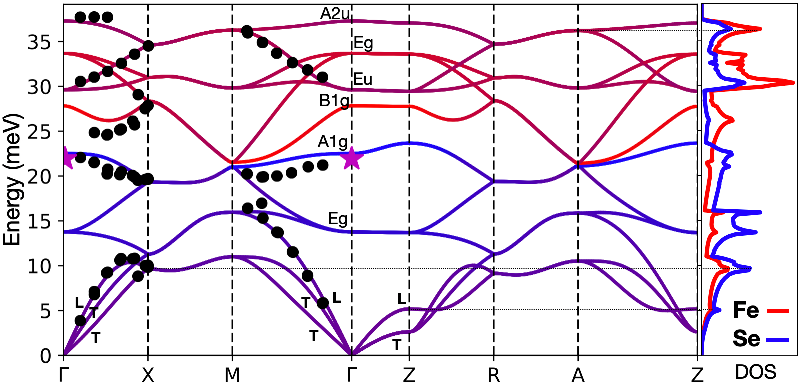}
	\caption{The phonon band dispersion in FeSe predicted by DFT+eDMFT. The black dots correspond to the inelastic x-ray scattering 
digitized from Ref.~\citenum{Murai_2020}, and the pink star is from X-ray data in Ref.~\citenum{Gerber_Science_2017}.}
	\label{fig_band}
\end{figure}
The phonon dispersion computed by DFT+eDMFT is displayed in Fig.~\ref{fig_band}. The black dots are digitized from experimental data of Ref.~\citenum{Murai_2020}, and the pink star corresponds to A1g phonon measurements in Ref.\citenum{Gerber_Science_2017}. At $\Gamma$ point we display the irreducible representations of all optical vibrations within the FeSe space group. We also mark the longitudinal (L) and transverse (T) acoustic modes in the figure.
The phonon branches along the high symmetry path in the Brillouin zone are colored according to their character of vibration. Predominantly iron (selenium) modes are red (blue).
On the right we also show the partial density of states of phonon vibrations with the same color scheme. The longitudinal and transverse branches are degenerate between $X=(1/2,0,0)$ and $M=(1/2,1/2,0)$ point, while the two transverse branches are degenerate between $\Gamma$ to $Z=(0,0,1/2)$ point. All acoustic modes are roughly equal mixture of Fe and Se vibrations, as their color is purple. We notice that all acoustic branches along high symmetry lines agree very well with the experiment of Ref.~\citenum{Murai_2020}.

\begin{figure}[t]
    \includegraphics[width=\columnwidth]{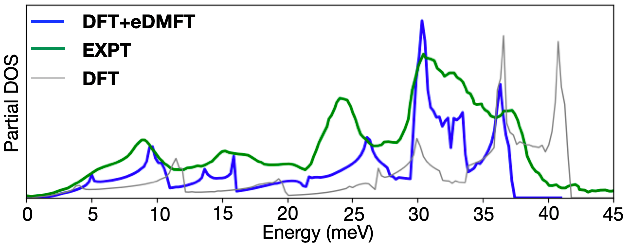}
    \caption{Fe partial density of states calculated with DFT and eDMFT compared with Fe nuclear inelastic scattering digitized from Ref.~\citenum{Cava_2010}.}
	\label{fig_dos}
\end{figure}
In Ref.~\citenum{Cava_2010} nuclear inelastic scattering was used to measure iron partial phonon-dos, which corresponds to our red curve on the right of Fig.~\ref{fig_band}. Fig.~\ref{fig_dos} shows again eDMFT and DFT calculated partial Fe-dos, and compares it to measurements from Ref.~\citenum{Cava_2010}. We notice a small peak around $5\,$meV in both eDMFT theory and experiment, which was assigned to the transverse mode around the $Z$ point in Ref.~\citenum{Cava_2010}. As can be seen in Fig.~\ref{fig_band} the theoretical transverse mode appears around $2.5\,$meV, but does not lead to any peak in the dos, while the longitudinal branches around $Z$ have frequency close to $5\,$meV and lead to first peak in the dos. DFT predicts the same peak at slightly lower energy, around $3\,$meV. This soft mode appears because the structure is highly anisotropic and layered with van der Walls inter-layer bonding along the $c$ axis, and large $c$-axis compressibility.

The second large peak around $9\,$meV was also assigned to transversal acoustic mode in Ref.~\citenum{Cava_2010}. Within eDMFT this mode is only a fraction of a meV higher than experiment (see Fig.~\ref{fig_dos}) and can be assigned to the saddle point around $X$ point of all three acoustic modes. Within DFT this mode appears at substantially higher energy around $12\,$meV. Hence, spin fluctuations stabilize the $c$-axis vibration and  increase the acoustic frequency at the $Z$ point, but make the in plane vibration at $X$ point softer, even though both modes are equal mixture of Fe and Se vibrations.

The next two peaks in Fe-dos appear around $14\,$meV and $17\,$meV. The first comes from the flat predominantly Se optical mode between $\Gamma-Z$, while the second comes from the top of acoustic modes at $M$ and $A$ point, also dominated by Se vibrations. The latter appears at slightly higher energy in experiment of Ref.~\citenum{Murai_2020} (less than $1\,$meV higher), as the black dots have slightly higher maximum. Comparing to Ref.~\citenum{Cava_2010}, we also see a broad hump of spectra between 15 and 18$\,$meV, which was assigned to longitudinal acoustic modes in Ref.~\citenum{Cava_2010}. While the largest peak around $17\,$meV is indeed from longitudinal vibrations around the $M$ point, the smaller peak in our calculation is due to optical mode. We notice that both peaks appear stronger in Se-partial dos, hence these vibrations involve strong Se-movement.

Next there is substantial dip in the dos between $21-22\,$meV both in theory and experiment, which comes from a small gap of the optical modes at $M$ point, where the lower and upper branch is predominantly of Fe and Se character. This is followed by a large peak, which appears around $24\,$meV in experiment, and $26\,$meV in eDMFT. This peak shows the largest discrepancy between eDMFT and experiment. However, we note that in DFT the same peak appears around $29\,$meV, similarly to all other optical modes, which appear at much higher in energy in DFT calculation. This peak comes from Fe-optical mode, which has a saddle point in-between $\Gamma-X$ point, and becomes B1g at $\Gamma$. Indeed the black dots corresponding to inelastic X-ray scattering~\cite{Murai_2020} around that point also show similar $2\,$meV downshift compared to eDMFT theory, hence we can assign the $24\,$meV peak to this in-plane optical mode half-way between $\Gamma-X$ of Fe-origin. Similar assignment was made in Ref.~\citenum{Cava_2010}.

Finally, the last three peaks between 30 and 40$\,$meV in the partial dos in Fig.~\ref{fig_dos} are in excellent agreement between experiment and eDMFT theory, therefore we can assign their origin. They are predominantly of iron origin. The largest 30$\,$meV peak comes from a very flat optical mode, which has Eu character at $\Gamma$-point. The saddle point and largest contribution to dos appears in the midpoint between $\Gamma$-$M$. Second, the shoulder around $34\,$meV comes from a flat Eg band, predominantly in the $\Gamma-Z$ direction. Third, the highest $37\,$meV peak comes from A2u optical mode, mostly from the saddle point at $M$ and $A$ point, while A2u mode at $\Gamma$ point is higher in energy and contributes to the upper tail of the peak. The assignment in Ref.~\citenum{Cava_2010} was different: The $30\,$meV, $34\,$meV and $37\,$meV were assigned as Eu, A2u, and Eg, respectively. On the basis of this theoretical calculation we can confidently claim that the right order should be Eu, Eg, A2u.

\begin{table}[t]
	\centering
	\begin{tabular}{c c c c} 
		\hline
		\hline
		Phonon Modes  & DFT & DFT+eDMFT & Experiment \\ 
		\hline
		\Eg & 19.4  & 13.8   & -  \\ 
		\Ag & 27.5  & 22.5   & $21.9{\cite{Gerber_Science_2017}}$, $21.8{\cite{JSNM}}$, $20.5{\cite{Zakeri_2017,PhysRevB.79.014519}}$, $20.6{\cite{Cava_2010}}$ \\ 
		\Bg & 30.5  & 27.7   & $27.1{\cite{JSNM}}$, $25.6{\cite{Zakeri_2017}}$, $24.5{\cite{PhysRevB.79.014519}}$, $25.5{\cite{Cava_2010}}$ \\ 
		\Eu & 37.2  & 29.7   & - \\ 
		\Eg & 39.4  & 33.6   & $35.1{\cite{JSNM}}$\\ 
		\Au & 42.0  & 37.3   & $38.7{\cite{Cava_2010}}$, $40{\cite{Zakeri_2017}}$, $39{\cite{PhysRevB.79.014519}}$ \\ 
		\hline
		\hline
	\end{tabular}
	\caption{\Gm point phonon frequencies of different phonon modes calculated by DFT and DFT+eDMFT compared with various experiments.}
	\label{table_freqs}
\end{table}
At the $\Gamma$ point we can decompose phonon modes into $\Ag+\Au+\Bg+Eu+2\Eg$ and we tabulated theoretical and experimental phonon frequencies in table~\ref{table_freqs}.
The DFT values are taken from the literature~\cite{Subedi_2008, Nakamura_2009, Cohen_2012, Xing_2014, Zakeri_2017}  and are consistent with our DFT calculation. The experimental values are determined by Raman scattering~\cite{JSNM}, inelastic neutron-scattering~\cite{PhysRevB.79.014519, Cava_2010}
electron energy-loss spectroscopy~\cite{Zakeri_2017}, and  time-domain X-ray scattering~\cite{Gerber_Science_2017}.
At the $\Gamma$ point all optical modes are softer in eDMFT and experiment compared to the DFT predictions, hence fluctuating moments tend to reduce the high-energy part of the phonon spectra. As pointed out above, this is not necessary the case for some acoustic modes. As can be seen in table~\ref{table_freqs}, eDMFT frequencies compare very well with experiments, and the disagreement between the two is at most $1.5\,$meV. On the other hand, the DFT values are considerably larger (up to $5\,$meV) deteriorating the agreement with experiments.

\begin{figure}[t]
	\includegraphics[width=\columnwidth]{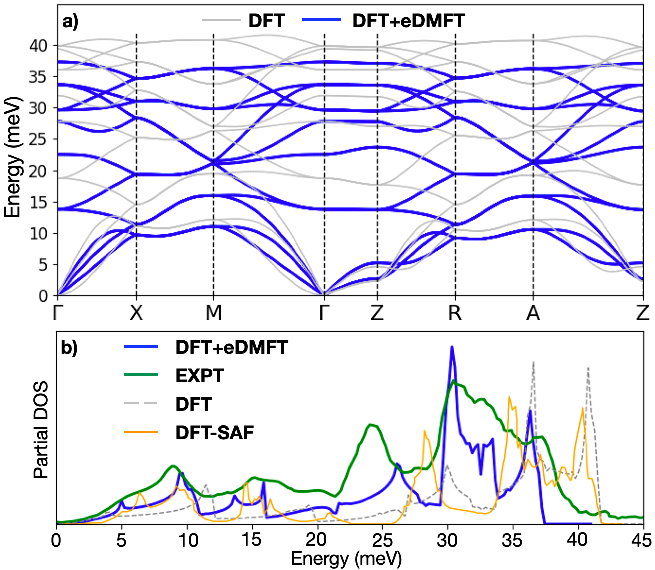}
    \caption{a) Calculated phonon band dispersion using DFT(PBE) and eDMFT. (b) Comparison of density of states between DFT stripe-antiferromagnetic state (SAF)  (see Ref.~\onlinecite{FeSe_AFM} for details), and experiment, eDMFT and non-magnetic DFT.}
    \label{fig_bands2}
\end{figure}
Finally, in Fig.~\ref{fig_bands2} we compare the results of magnetic and non-magnetic DFT with the eDMFT and experimental results. Fig.~\ref{fig_bands2}(a) shows dispersion within non-magnetic DFT (GGA) and eDMFT. We notice that the topology of bands is similar in the two theories, however, 
most of optical modes are shifted to higher energy in DFT theory, in particular the first two optical modes (Eg and A1g) which are shifted for almost $5\,$meV throughout the Brillouin zone. We notice that these are predominantly of Se character. However, even the higher energy modes, which are predominantly of Fe character, shift to higher frequency for a few meV.
Contrary to most optical modes, the transverse acoustic modes are at similar frequency, and the longitudinal mode around $Z$ point is even shifted to lower energy, as pointed out above.
The fluctuating moment on iron thus not only affects the iron modes, but surprisingly, it affects the predominantly Se modes even stronger. This is somewhat counterintuitive, as local correlations on Fe atom, as included in eDMFT, would be naively expected to modify Fe vibrations more than Se.

To understand if static magnetic moments incorporated in magnetic DFT calculation can mimic fluctuating moments, we compare in Fig.~\ref{fig_bands2}(b) the results of the stripe-antiferromagnetic state (orange curve) with DFT, eDMFT results and experiment. This stripe state is the competing magnetic state in most iron superconductors. We notice that some peak positions are greatly improved in magnetic DFT as compared to non-magnetic DFT, for example the $9\,$meV acoustic peak is correctly softened, and appears very close to experimental position. On the other hand, the $5\,$meV peak is hardened, and appears at too high energy (around $7\,$meV). The acoustic modes between $20\,$meV and $40\,$meV all appear at much too high frequency, similar to DFT results, and are barely changed from DFT. The only exception is the Fe-optical mode from the saddle point between $\Gamma-X$. This peak appears in experiment at around $24\,$meV, in DFT around $30\,$meV and in spin-polarized DFT around $28\,$meV, and finally in eDMFT around $26\,$meV. Hence, with the exception of a few acoustic modes, the majority of the vibrations are not much better predicted by static magnetic calculation as compared to non-magnetic DFT. Hence, the fluctuating magnetic moments in such metallic environment show very different phonon spectra than static analogs, even though the equilibrium structures are very similar.

{\color{NavyBlue}\textit{Summary:}} In summary, proper description of fluctuating moments on Fe atoms in FeSe, as described by DFT+eDMFT theory, gives phonon dispersion and phonon density of states in very good agreement with experiments, and largely eliminates the deficiency of DFT. The inclusion of static magnetic moments on Fe, as opposed to dynamic moments, does not appreciably improve DFT results, because the electronic band structure in these metallic system is not correctly predicted within theories with static moments. We also show that vibrations with mostly Se-character are modified even stronger than those of Fe-character, even though fluctuating moments appear on Fe atoms only.
The electronic correlations in metallic systems have a major effect on lattice dynamics, and proper inclusion of fluctuating moments local to the correlated atoms, as described by Dynamical Mean Field Theory, substantially improves accuracy of the lattice dynamics. This will be important in database and data science driven material design, which is currently based mostly on DFT engine, but has recently been integrated with DMFT as well~\cite{Mandal2019,choudhary2020jarvis}.

{\color{NavyBlue}\textit{Acknowledgments:}}
This work was supported by NSF DMR-1709229. GK is grateful to Subhasish Mandal and Gheorghe Lucian Pascut for many fruitful discussion.

\bibliographystyle{apsrev4-1} 
\bibliography{FeSe}

\end{document}